\documentclass[aps,prl,showpacs,twocolumn,groupedaddress,superscriptaddress]{revtex4}
\usepackage{graphicx}%
\usepackage{amsmath}
\usepackage{dcolumn}% Align table columns on decimal

\begin{document}

\title{Routes to synchrony between asymmetrically interacting oscillator ensembles}

\author{Jane H.~Sheeba}%
\affiliation{Department of Physics, Lancaster University,
Lancaster, LA1 4YB, UK}

\author{V.~K.~Chandrasekar}%
\affiliation{Department of Physics, Lancaster University,
Lancaster, LA1 4YB, UK}

\author{Aneta~Stefanovska}%
\affiliation{Department of Physics, Lancaster University,
Lancaster, LA1 4YB, UK}
\affiliation{Faculty of Electrical Engineering, University of Ljubljana, Tr$\breve{z}$a$\breve{s}$ka
25, 1000 Ljubljana, Slovenia}

\author{Peter~V.~E.~McClintock}%
\affiliation{Department of Physics, Lancaster University,
Lancaster, LA1 4YB, UK}

\date{\today}

\begin{abstract}
We report that asymmetrically interacting ensembles of
oscillators (AIEOs) follow novel routes to synchrony. These
routes seem to be a characteristic feature of coupling
asymmetry. We show that they are unaffected by white noise
except that the entrainment frequencies are shifted. The
probability of occurrence of the routes is determined by phase
asymmetry. The identification of these phenomena offers new
insight into synchrony between oscillator ensembles and suggest
new ways in which it may be controlled.
\end{abstract}



\maketitle

Coupling asymmetry between oscillators is widespread in real
physical and biological systems. Examples include
cardio-respiratory and cardio-$\delta$(EEG) interactions
\cite{Stefanovska:07}, interactions among activator-inhibitor
systems \cite{Daido:06}, coupled circadian oscillators
\cite{Fukuda:07}, and the interactions between ensembles of
oscillators in neuronal dynamics \cite{Sherman:92,Roelfsema:97}.
Synchrony can be sometimes be desirable as in e.g.\ lasers and
Josephson-Junction arrays \cite{Trees:05}, or temporal coding and
cognition via brain waves \cite{Singer:99}; but it can also be
dangerous, as in epileptic seizures \cite{Timmermann:03},
Parkinson's tremor \cite{Percha:05}, or pedestrians on the
Millennium Bridge \cite{Strogatz:01}. Thus the control of
synchronization \cite{Blasius:03} can often be important. An
understanding how synchrony arises is of course an essential
prerequisite for the development of control schemes.

In this Letter, we report a phenomenon that occurs in
asymmetrically interacting ensembles of oscillators (AIEOs), where
it provides novel routes to synchrony. These routes are the characteristic
of AIEOs and they cannot occur
in systems where the interactions are symmetrical. Our results
yield new insights into how synchrony arises and offer possible
ways of controlling synchrony between real AIEOs.

The phase dynamical equations of a system of two AIEOs can be
written as \cite{Kuramoto:84}
\begin{eqnarray}
\label{mod01}
 \dot{\theta_i}^{(1,2)}&=& \omega_i^{(1,2)} - \frac{A^{(1,2)}}{N^{(1,2)}}\sum_{j=1}^{N^{(1,2)}}
\sin(\theta_i^{(1,2)}-\theta_j^{(1,2)}+\alpha^{(1,2)}) \nonumber \\
 &&\quad -
 \frac{B}{N^{(2,1)}}\sum_{j=1}^{N^{(2,1)}}\sin(\theta_i^{(1,2)}-\theta_j^{(2,1)}+\alpha^3).
\end{eqnarray}
The interactions are characterized by coupling parameters
$A^{(1,2)}$ and $B$ to quantify respectively the interactions
within, and between, the ensembles. The fact that $A^{(1)} \neq
A^{(2)}$ (c--asymmetry hereafter) implies that the oscillators
in the ensembles are asymmetrically coupled. $\theta_i^{(1,2)}$
are the phases of the $i$th oscillator in each ensemble and
$N^{(1,2)}$ refer to the ensemble sizes; we take
$N^{(1)}=N^{(2)}=N$, with $N\rightarrow\infty$. Phase asymmetry
(p--asymmetry hereafter) is introduced by phase shifts
$0\leq\alpha^{(1,2,3)}<\pi/2$. The natural oscillator
frequencies $\omega_i^{(1,2)}$ are assumed to be Lorentzianly
distributed as $g(\omega^{(1,2)}) =
\frac{\gamma}{\pi}(\gamma^2+(\omega-\bar{\omega}^{(1,2)})^2)^{-1}$
with central frequencies $\bar\omega^{1,2}$, and $\gamma$ is
the half--width at half--maximum. With this characterization,
we show that an increase of the coupling strength between two
ensembles that are synchronized separately does not immediately
result in their mutual phase-locking. Rather, phase-locking
occurs through either one of two different routes: in Route-I
the oscillators in the two ensembles combine and form clusters;
in Route-II one of the ensembles desynchronizes while the other
remains synchronized. Further, there also exists the
possibility that phase-locking between the ensembles cannot
occur at all.

The model can conveniently be expressed in terms of
(complex-valued, mean-field) order parameters
$r^{(1,2)}e^{i\psi^{(1,2)}}=\frac{1}{N}\sum_{j=1}^N
e^{i\theta_j^{(1,2)}}$. Here $\psi^{(1,2)}(t)$ are the average
phases of the oscillators in the respective ensembles and
$r^{(1,2)}(t)$ provide measures of the coherence of each
oscillator ensemble which varies from 0 to 1. When
$r^{(1,2)}\approx 1$ the corresponding ensemble is synchronized in
phase (microscopic synchronization) and when
$\delta\psi=\psi^{(1)}-\psi^{(2)}\approx$ constant the ensembles are
mutually locked in phase (macroscopic synchronization). With these
definitions, Eq.\ (\ref{mod01}) becomes
\begin{eqnarray}
\label{mod04} \dot{\theta_i}^{(1,2)}&=&\omega_i^{(1,2)}-A^{(1,2)} r^{(1,2)}\sin(\theta_i^{(1,2)}-\psi^{(1,2)}+\alpha) \nonumber \\
&&\quad - Br^{(2,1)}\sin(\theta_i^{(1,2)}-\psi^{(2,1)}+\alpha),
\end{eqnarray}
where for simplicity we have considered the particular case
$\alpha^{(1,2,3)}=\alpha$. In the limit $N\rightarrow \infty$, a
density function can be defined as
$\rho^{(1,2)}(\theta,t,\omega)d\omega d\theta$ which describes the
number of oscillators with natural frequencies within
$[\omega,\omega+d\omega]$ and with phases within
$[\theta,\theta+d\theta]$ at time $t$. For fixed $\omega$ the
distribution $\rho^{(1,2)}(\theta,t,\omega)$ obeys the evolution
equation $\partial\rho^{(1,2)}/\partial
t=-\partial(\rho^{(1,2)}\dot{\theta}^{(1,2)})/\partial\theta$. The
function $\rho^{(1,2)}(\theta,t,\omega)$ is real and $2\pi$
periodic in $\theta$, so it can be expressed as a Fourier series
in $\theta$
\begin{eqnarray*}
\rho^{(1,2)}(\theta,t,\omega)&=&\sum_{l=-\infty}^{\infty}\rho_l^{(1,2)}(\omega,t)e^{il\theta}\\
&=&\frac{1}{2\pi}+\rho_1^{(1,2)}e^{i\theta}+\mbox{c.c}+\eta(\theta,t,\omega),
\end{eqnarray*}
where c.c is the complex conjugate of the preceding term and
$\eta(\theta,t,\omega)$ denotes the 2nd and higher harmonics.
Substituting $\rho^{(1,2)}(\theta,t,\omega)$ into the evolution
equation, we get the following linearized equation for $\rho_1$ as
\begin{eqnarray}
\label{any05}
 \dot{\rho_1}^{(1,2)}&+&i\omega \rho_1^{(1,2)}=
\frac{e^{i\alpha}}{2}\left(A^{(1,2)}\langle \rho_1^{(1,2)}\rangle+B\langle \rho_1^{(2,1)}\rangle\right)
,
\end{eqnarray}

\begin{figure}[t!]
%\begin{center}
\includegraphics[height=4cm,width=8.5cm]{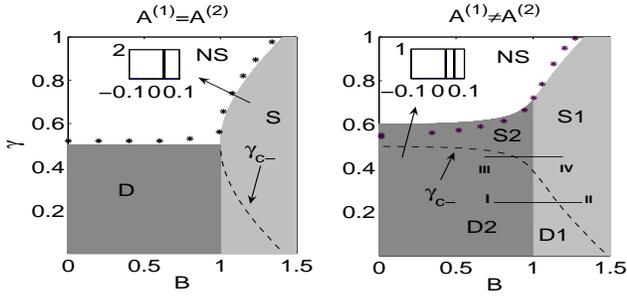}
\caption{($B$--$\gamma$) bifurcation diagram for
$\alpha=0,\Delta\omega=1$. The different synchronization regimes
are described in the text. The boundary between regimes NS and S/D
represents $\gamma_{c+}$. The $*$s represent the numerical
bifurcation boundaries; (left) $A^{(1)}=A^{(2)}=1$, (right)
$A^{(1)}=1.2,A^{(2)}=1$. Insets show the frequency distributions (also
obtained numerically) for the indicated regions; their ordinate
axes represent oscillator counts in thousands. Note that the
occurrence of perfect synchronization with $2000$ (left) and
$1000$ (right) oscillator groups will not occur throughout all of
each indicated region.} \label{gmp1}
%\end{center}
\end{figure}

\noindent where the Fourier components for $|l|\geq2$ are neglected since
$l=\pm1$ are the only nontrivial unstable modes and $\rho_0=1/2\pi$
is the trivial solution corresponding to incoherence.
$\langle\cdot\rangle$ represents the average over the frequencies
$\omega^{(1,2)}$ weighted by the Lorentzian distribution $g(\omega^{(1,2)})$.
The eigenvalues
obtained from the characteristic equation of (\ref{any05}) are
\begin{eqnarray}
\lambda_{\pm} = \left\{
\begin{array}{ll}
-\gamma+\frac{\kappa}{4}e^{i\alpha}\pm\frac{1}{2}(p^2+q^2)^{\frac{1}{4}}e^{i\frac{1}{2}\zeta}
-i\bar{\omega},&\\ \qquad\qquad\qquad\qquad\qquad\qquad\qquad p>0&
\\\\
-\gamma+\frac{\kappa}{4}e^{i\alpha}\mp\frac{i}{2}(p^2+q^2)^{\frac{1}{4}}e^{i\frac{1}{2}\zeta}
-i\bar{\omega},&\\ \qquad\qquad\qquad\qquad\qquad\qquad\qquad p<0&
\end{array}
\right. \label{any06}
\end{eqnarray}
where $\kappa=A^{(1)}+A^{(2)}$, $\xi=(\frac{1}{4}\hat{A}^2+B^2)$,
$\hat{A}=(A^{(1)}-A^{(2)})$, $\zeta=\tan^{-1}(\frac{q}{p})$,
$\Delta\omega=\bar{\omega}^{(1)}-\bar{\omega}^{(2)}$,
$\bar{\omega}=(\bar{\omega}^{(1)}+\bar{\omega}^{(2)})/2$,
%\begin{eqnarray*}
$p=\xi\cos(2\alpha)+\hat{A}\Delta\omega\sin{\alpha}-\Delta\omega^2$,
{\rm and} $q=\xi\sin(2\alpha)-\hat{A}\Delta\omega\cos{\alpha}$.
%\end{eqnarray*}
As a signature of synchronization, we take the condition
$Re(\lambda_{\pm})>0$ for analytic treatment. For the numerical
experiment, we set $r^{(1,2)}>0.7$ for microscopic synchronization
in the corresponding ensembles and a constant $\delta\psi$ for
macroscopic synchronization as the conditions. We take $N=1000$ in
each ensemble and the equations are solved using a fourth-order
Runge--Kutta routine. The initial phases of the oscillators are
assumed to be equally distributed in $[0,2\pi)$.

One might intuitively anticipate the possibility of four distinct
dynamical regimes: no synchronization (NS); global synchronization,
in which the oscillators of both ensembles are entrained to the same
frequency (S1); synchronization within one ensemble but not the
other (S2); synchronization within both ensembles, separately and
independently, with two entrainment frequencies (D2). In what
follows, we show that this is indeed the case and, furthermore, that
there is a global regime in which the
two ensembles behave as one, but oscillators within each ensemble
are entrained at either one of two distinct entrainment frequencies
(D1). Regions S2 and D1 cannot occur when c--asymmetry is absent
\cite{Okuda:91} (see Fig. \ref{gmp1}).

\begin{figure}[t!]
%\begin{center}
\includegraphics[height=3.7cm,width=8.5cm]{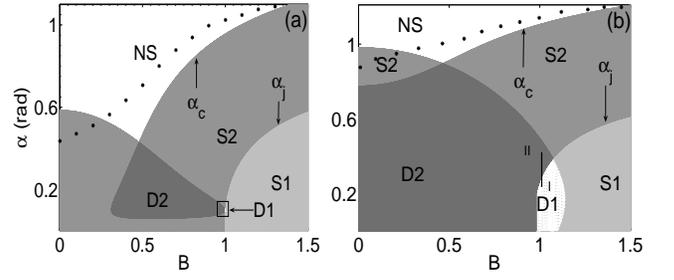}
\caption{($B$--$\alpha$) bifurcation diagram for (a)
$A^{(1)}=1.2,A^{(2)}=1$, (b) $A^{(1)}=1.8,A^{(2)}=1.4$ and
$\Delta\omega=1,\gamma=0.5$. Note that the synchronization regimes
S1 and D1 are greatly reduced in the presence of p--asymmetry. The
$*$s denote the numerical bifurcation boundary between the
synchronized and incoherent states. The discrepancy between
numerical and analytical boundaries occurs due the p--asymmetry
(unlike Fig. \ref{gmp1}) that affects region S2 and hence the
thresholds for $r^{(1)}$ and $r^{(2)}$.} \label{gmp2}
%\end{center}
\end{figure}

\begin{figure}[t!]
%\begin{center}
\includegraphics[height=6cm,width=7cm]{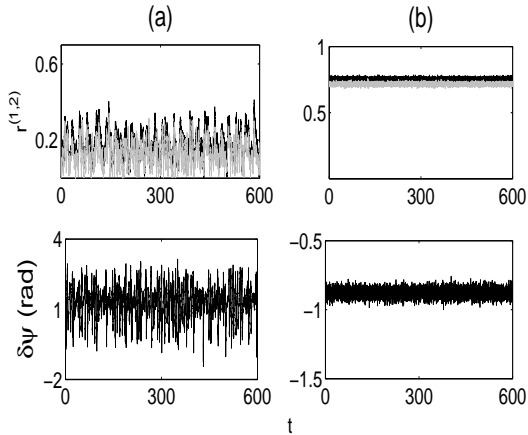}
\caption{Numerical plot of coherence parameters $r^{(1)}$ (grey),
$r^{(2)}$ (black) and phase difference $\delta\psi$ as a function of
time. Here (a) $B=1$, $\alpha=0.23$ and (b) $B=1$, $\alpha=0.47$.
(a) and (b) correspond to regions D1 and D2 of Fig. \ref{gmp2}(b)
for the same values of parameters as traveling along the line
$I-II$. Note that the order parameters display no synchronization
in the D1 region.}\label{gmp5}
%\end{center}
\end{figure}

\begin{figure}[t!]
%\begin{center}
\includegraphics[height=3cm,width=7cm]{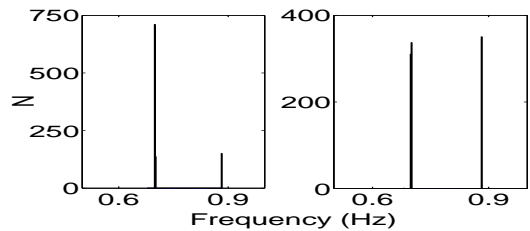}
\caption{Histograms showing virtual desynchronization - Left:
First ensemble, Right: Second ensemble, for the corresponding
parameter values in Fig. \ref{gmp5} (a). Note that the first
frequency component in both the ensembles have two
indistinguishably different subcomponents which may be considered
as one. This occurs due to the discrepancy between numerics and
analytics.}\label{gmp7}
%\end{center}
\end{figure}

For the case $\alpha=0$, when $p>0$,
in the region S1, the incoherent (steady) state becomes
unstable via a single Hopf bifurcation and the ensembles
entrain to a single frequency $\Omega_+$. With further decrease
of $\gamma$ below the $\gamma_{c-}$ line in the region D1, a
new entrainment frequency emerges through a second Hopf
bifurcation. In this region, the oscillators from the two
ensembles combine and form two clusters (macroscopic
clustering) oscillating with two frequencies,
$\Omega_{\pm}=-\mbox{Im}(\lambda_{\pm})=\pm (1/2)[p^2
+q^2]^{\frac{1}{4}} \sin{(\frac{1}{2}\zeta)}+\bar{\omega}$.
$\gamma_{c\pm}$ are obtained by imposing the condition
$Re(\lambda_{\pm})=0$. Thus in this region the order parameters
either fluctuate in a quasi--periodic manner or have
complicated dynamics (see Fig. \ref{gmp5}). This is because
each ensemble has two clusters oscillating with different
frequencies (see Figs. \ref{gmp5}(a) and \ref{gmp7}) \cite{01}.
The presence of two entrainment frequencies can be seen by
looking at the frequencies into which all the individual
oscillators are grouped as shown in Fig. \ref{gmp7}. The
macroscopic clustering that occurs in this case is quite
different from formation of clusters in a single ensemble
\cite{Kuramoto:84,Golomb} - here the oscillators in two
different ensembles \emph{combine} and form clusters. The
occurrence of this phenomenon provides new insight into the
control of synchrony in realistic situations where there is
asymmetry, like neural networks where neurons from one ensemble
(e.g.\ cortex) tend to synchronize with those in the other
ensemble (e.g.\ thalamus) thus giving rise to creating
desirable or undesirable (e.g.\ epileptic seizures) effects. In
the absence of coupling asymmetry, these phenomena do not
exist. Traversing the line I--II of Fig. \ref{gmp1}
demonstrates Route--I to phase-locking of the ensembles. As we
increase $B$, the oscillators in the ensembles pass from the
dynamical state of microscopic synchronization (D2) through
macroscopic clustering to macroscopic synchronization or
phase-locking of the ensembles. Note that when
$A^{(1)}=A^{(2)}$ or $\Delta\omega=0$ only one entrainment
frequency exists below $\gamma_{c-}$ and therefore this
sequence does not occur (due to the absence of region D1).
While traversing the line III-IV of Fig. \ref{gmp1} the
ensembles adopt Route-II to phase-locking in which it is the
macroscopic clustering that does not occur.

When $p<0$, corresponding to regions S2 and D2 in Fig.\
\ref{gmp1}, the dynamics is the same as above except that the
values of critical $\Gamma$ and synchronization frequencies
differ, as can be calculated from Eq.\ (\ref{any06}). In region
S2, microscopic synchronization can occur in either one of the
ensembles, depending upon whether $A^{(1)}$ or $A^{(2)}$ is
greater; in Fig. \ref{gmp1}, since $A^{(1)}>A^{(2)}$,
synchronization occurs in the first ensemble with the second
ensemble remaining incoherent. Note that, on increasing $B$ while in region S2,
the condition $p<0$ is violated and the ensembles enter into
the phase-locked region S1. In region D2, the ensembles
synchronize separately to two locking frequencies (unlike in
region D1 where the ensembles combine). The corresponding
$(B-\gamma)$ bifurcation diagram for the case
$A^{(1)}$=$A^{(2)}$ is plotted in Fig. \ref{gmp1} (left) to
show the difference between these two cases. The region D
represents microscopic synchronization which occurs through a
degenerate Hopf bifurcation (similar to D2) and S represents
macroscopic synchronization through a single Hopf bifurcation
(similar to S1). The regions S2 and D1 cannot arise for this
case. Hence the ensembles cannot follow either Route-I or
Route-II to synchrony but can only pass from the dynamical
states of microscopic synchrony to macroscopic synchrony.

\begin{figure}[b!]
%\begin{center}
\includegraphics[height=6cm,width=7cm]{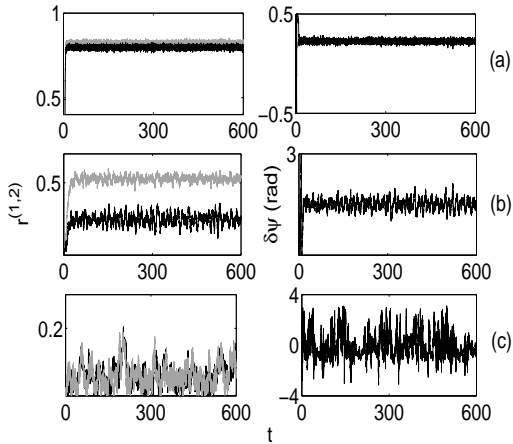}
\caption{Numerical plot of coherence parameters $r^{(1)}$ (grey), $r^{(2)}$
(black) and phase difference $\delta\psi$ as a function of time. (a)
$B=0.7, \alpha=0.2$ (b) $B=0.7, \alpha=\pi/4$ (near $\alpha_c$) (c)
$B=0.7, \alpha=1.2$ ($\alpha>\alpha_c$). (a), (b) and (c) correspond
to regions D2, near NS/S2 boundary and NS of Fig. \ref{gmp2}(a) for the same values
of parameters. }\label{gmp3}
%\end{center}
\end{figure}

For the case $\alpha\ne0$, regions D1 and S1 shrink as $\alpha$
increases, whereas S2 expands, as shown in Fig. \ref{gmp2}.
This means that p--asymmetry reduces the probability of
macroscopic synchronization (reduced S1 and D1 regions in Fig.
\ref{gmp2}) and mostly allows only microscopic synchronization
of one or both of the ensembles. For a given set of parameters,
there exists a value of $\alpha$ below which the condition
$p>0$ is satisfied and above which $p<0$ is satisfied. As a
result, when $\alpha>\alpha_j$ the macroscopic synchrony breaks
and the system enters the microscopically synchronized state.
Thus as one travels from S1 (D1) to S2 (D2) across $\alpha_j$
the combined synchrony with single (double) frequency breaks
between the ensembles and independent synchronization with
single (double) frequency regime appears. Region S2, unlike in
Fig. \ref{gmp1}, embraces two states (i) synchronization in
ensemble 1 with ensemble 2 incoherent and (ii) synchronization
in ensemble 2 with ensemble 1 incoherent, but does distinguish
between them. Further, there is a critical value of
$\alpha=\alpha_c$ above which the collective oscillations
disappear and the incoherent state becomes stabilized (see
Figs. \ref{gmp2} and \ref{gmp3}).

\begin{figure}[t!]
%\begin{center}
\includegraphics[height=2.5cm,width=7cm]{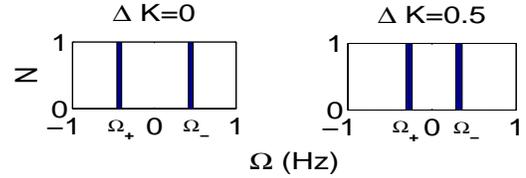}
\caption{The shift in the entrainment frequencies $\Omega_{\pm}$
with $\Delta K$. Here $A^{(1)}=A^{(2)}$, $B=0.5$, $\alpha=0$, $\bar\omega=0$
and $\Delta\omega=1$. $N$ represents oscillator count in thousands.
}\label{gmp4}
%\end{center}
\end{figure}

Real physical systems are of course subject to noise (random
fluctuations, of either internal or external origin), so we now
consider how the above analysis will be modified as a result. Adding
$\eta_i^{(1,2)}$
to the RHS of Eq.\ (\ref{mod01}), where $\eta_i^{(1,2)}$ are independent
Gaussian white noises with
$\langle\eta_i^{(1,2)}(t)\rangle =0$ and
$\langle\eta_i^{(1,2)}(t)\eta_j^{(1,2)'}(t)\rangle
=2K^{(1,2)}\delta(t-t')\delta_{ij}$ and $K^{(1,2)}$ are the noise
intensities, the eigenvalues of the linearized equation then take
the form
\begin{eqnarray}
\lambda_{\pm} = \left\{
\begin{array}{ll}
-\bar{K}-\gamma+\frac{\kappa}{4}e^{i\alpha}\pm\frac{1}{2}(p^2+q^2)^{\frac{1}{4}}e^{i\frac{1}{2}\zeta}
-i\bar{\omega},&\\ \qquad\qquad\qquad\qquad\qquad\qquad\qquad p>0&
\\\\
-\bar{K}-\gamma+\frac{\kappa}{4}e^{i\alpha}\pm\frac{i}{2}(p^2+q^2)^{\frac{1}{4}}e^{i\frac{1}{2}\zeta}
-i\bar{\omega},&\\ \qquad\qquad\qquad\qquad\qquad\qquad\qquad p<0&
\end{array}
\right. \label{anyn1}
\end{eqnarray}
where $\bar{K}=\frac{(K^{(1)}+K^{(2)})}{2}$, $\Delta
K=K^{(1)}-K^{(2)}$
$p=\xi\cos(2\alpha)+\hat{A}[\Delta\omega\sin{\alpha}+\Delta K
\cos(\alpha)]-\Delta\omega^2+\Delta K^2$ and
$q=\xi\sin(2\alpha)-\hat{A}\Delta\omega\cos{\alpha}-\hat{A}\Delta
K\sin{\alpha}+2\Delta\omega\Delta K$. For simplicity, if we
consider $K^{(1)}=K^{(2)}=K$, one can then replace $\gamma$ in
Eq. (\ref{any06}) by $\gamma+K$. Thus from Eq. (\ref{anyn1}) it
is evident that the dynamics and hence the new route to
synchrony are unaffected by the presence of white noise.
Consequently, for increasing noise intensity, the incoherent
state becomes unstable for larger values of the critical
parameters. When $K^{(1)}\neq K^{(2)}$ the entrainment
frequency depends on $\Delta K$ (shown in Fig. \ref{gmp4}). On
increasing $\Delta K$ ($K^{(1)}>K^{(2)}$), the microscopic
synchronization of ensemble 1 is destroyed.

In summary, we have found new routes to synchrony between two
AIEOs, characterized by asymmetry in the coupling strength. We
have also established that the effect of white noise is simply
to alter the critical values of the parameters that control
bifurcation, and to change the entrainment frequencies, but
without otherwise affecting this route to synchrony. Our main
focus has been on phase oscillator ensembles, but similar
phenomena can also be identified in ensembles of other kinds of
oscillators.

The authors are indebted to A. Bahraminasab, D. Garcia-Alvarez and
M. G. Rosenblum for valuable discussions. The study was supported by
the EC FP6 NEST-Pathfinder project BRACCIA and in part by the
Slovenian Research Agency and the Council of Scientific and
Industrial Research, India.


\begin{thebibliography}{10}
\bibitem{01}
Even though Figs. \ref{gmp5} and \ref{gmp7} are plotted for the
case $\alpha\neq0$, they still describe quite well cases where
$\alpha=0$. This is because finite $\alpha$ does not change the
dynamical behavior in region D1, though it does changes the size
of this region.
\end{thebibliography}
\end{document}